\documentclass[review]{elsarticle}
\usepackage{hyperref}
\usepackage{float}
\usepackage[table]{xcolor}
\usepackage{lipsum}
\usepackage{multirow}
\usepackage{setspace}
\usepackage{caption}
\usepackage{graphics}
\usepackage{graphicx}
\usepackage{color} \usepackage{numcompress}
\usepackage{amsmath, subfigure}
\usepackage{amssymb}
\usepackage{amsfonts}
\usepackage[utf8]{inputenc}
\usepackage[english]{babel}
\usepackage{amsthm}
\theoremstyle{definition}

\theoremstyle{theorem}
\newtheorem{theorem}{Theorem}[section]
\theoremstyle{lemma}
\newtheorem{lemma}{Lemma}[section]
\usepackage{multirow,enumerate}
\usepackage{epstopdf}
\usepackage{epsfig}
\usepackage{bm}
\usepackage{booktabs,multirow,tabularx}

\definecolor{brcolor}{rgb}{0.8,0.0,0.4}

\newcommand{\be}{\begin{eqnarray}}

\newcommand{\ee}{\end{eqnarray}}           \newcommand{\ba}{\begin{eqnarray*}}
\newcommand{\ea}{\end{eqnarray*}}

\newcommand{\bt}{\mathbf{t}}

\journal{Journal of \LaTeX\ Templates}

\usepackage{numcompress}\biboptions{authoryear}

\begin{document}

\begin{frontmatter}

\title{MEDL and MEDLA: Methods for Assessment of Scaling by Medians of Log-Squared Nondecimated Wavelet Coefficients}

\author{Minkyoung Kang and Brani Vidakovic} 
\address{ H. Milton Stewart School of Industrial \& Systems Engineering, Georgia Institute of Technology, 765 Ferst Drive NW, Atlanta, Georgia 30332}

%
%

\begin{abstract}
High-frequency measurements and images acquired from various sources in the real world often possess a degree of self-similarity and inherent regular scaling. When data look like a noise,
 the scaling exponent may be the only informative feature that summarizes such data. Methods for the assessment of self-similarity by estimating Hurst exponent 
often involve analysis of rate of decay in a spectrum defined in various multiresolution domains. When this spectrum is calculated
 using discrete non-decimated wavelet transforms, due to increased autocorrelation in wavelet coefficients, the estimators of $H$ show  increased bias compared to the estimators that use traditional orthogonal transforms.
At the same time, non-decimated transforms have a number of advantages when employed for calculation of  wavelet spectra and estimation of Hurst exponents: the variance of the estimator is smaller, input signals and images could be of arbitrary size, and due to the shift-invariance, the local scaling can be assessed as well.
We propose two methods based on robust estimation and resampling that alleviate the effect of increased autocorrelation  while maintaining all advantages of non-decimated wavelet transforms. The proposed methods extend the approaches in existing literature where the logarithmic transformation and pairing of wavelet coefficients are used for lowering the bias.

  In a simulation study we use fractional Brownian motions with a range of theoretical Hurst exponents.
   For such signals for which ``true'' $H$ is known, we demonstrate bias reduction and overall
  reduction of the mean-squared error by the two proposed estimators.
  For fractional Brownian motions, both proposed methods yield
  estimators of $H$ that are asymptotically normal and unbiased.

 \end{abstract}

\begin{keyword}
Non-decimated wavelet transform, scaling, and Hurst exponent
\end{keyword}

\end{frontmatter}


\section{Introduction}

At first glance, data that scale look like noisy observations, and often
the large scale features
(basic descriptive statistics, trends, smoothed functional estimates, etc.)
carry no useful information.
For example, the pupil diameter in humans fluctuates at a high frequency
(hundreds of Hz), and prolonged monitoring leads to
massive data sets. Researchers found that the high-frequency dynamic of change
in the diameter is informative of eye pathologies, e.g., macular degeneration, \cite{moloney}.
Yet, the trends and traditional summaries of the data
are clinically irrelevant, for the magnitude of the diameter depends on the ambient light,
and not on the inherent eye pathology.

Our interest focuses on the analysis of self-similar objects.
Formally, a deterministic function $f(\bt)$ of a $d$-dimensional argument $\bt$ is said to be
self-similar  if $f(\lambda \bt)=\lambda^{-H}f(\lambda \bt), $ for some choice of the exponent
$H$, and for all dilation factors $\lambda$.  The notion of
self-similarity has been extended to random processes. Specifically, a
stochastic process $\{X(\bt),\ \bt\in R^d\}$ is self-similar with
scaling exponent (or \emph{Hurst exponent}) $H$ if, for any
$\lambda \in R^+$,
\begin{eqnarray}\label{basicdef}
X(\lambda \bt)\stackrel{d}{=}\lambda^H X(\bt),
\end{eqnarray}
where the relation ``$\stackrel{d}{=}$'' is understood as the equality in all finite dimensional distributions.

In this paper, we are concerned with a precise estimation
 of scaling exponent $H$ in one-dimensional setting. The results can be readily extended to self-similar objects of arbitrary number of dimensions. \\

 A number of estimation methods for $H$ exist, including:
  re-scaled range calculation ($R/S$), Fourier-spectra methods,
variance plots, quadratic variations, zero-level crossings, etc.
For a comprehensive
description, see \cite{beran1994}, \cite{doukhan2003theory}, and \cite{abry2013}.  Wavelet transforms are
especially suitable for modeling self-similar phenomena,  as is reflected by vibrant research.
An overview is provided in \cite{abry2000b}.

If processes possess a stochastic structure (e.g. Gaussianity, stationary increments),
the scaling exponent $H$ becomes a parameter in a well-defined statistical model
and can be estimated as such.
Fractional Brownian motion (fBm)
is important and
well understood model for data that scale. Its importance follows from the fact that
fBm is a unique Gaussian process  with stationary increments that
is self-similar, in the sense of (\ref{basicdef}).

A fBm has a
(pseudo)-spectrum of the form $S(\omega)\propto
 |\omega|^{-(2H+1)}$, and consequently the log-magnitudes
of detail coefficients at different resolutions in a wavelet decomposition exhibit a linear
relationship. Using non-decimated wavelet domains to leverage on this linearity constitutes the staple of this paper.

Each decomposition level in nondecimated wavelet transform (NDWT) contains the same number 
of coefficients as the size of the original signal. This redundancy contributes to the accuracy of estimators of $H$. However, reducing the bias induced by level-wise correlation among the redundant
coefficients becomes an important issue.
The two estimators we propose are based on the so-called ``logarithm-first'' approach, connecting Hurst exponent with a robust location
and resampling techniques.

The rest of the paper consists of three additional sections and an appendix. Section 2 provides background of wavelet transforms as well as the properties of resulting wavelet coefficients. Section 3 presents distributional results on which the proposed methods are based. Section 4 provides the simulation results and compares the estimation performance of the proposed methods to some standardly used methods. The final Section is reserved for concluding remarks.  Appendix contains all technical details for the results
  presented in Section 3.

\section{Orthogonal and non-decimated wavelet transforms}
Discrete signals from an acquisition domain can be mapped to the wavelet domain in multiple ways. We overview two versions of discrete wavelet transform: orthogonal wavelet transform (DWT) and non-decimated wavelet transform (NDWT). We also describe algorithmic procedures in performing two versions of wavelet transform and obtaining the wavelet coefficients. Here we focus on functional representations of wavelet transform which is more critical for the subsequent derivations. Interested readers can refer to \cite{nason1995}, \cite{vidakovic1999}, and \cite{percival2006wavelet} for alternative definitions.\\

Any square-integrable $L_2(\mathbb{R})$ function $f(x) $ can be represented in the wavelet domain as
\ba
f(x)=\sum_{k}c_{J_0, k}\phi_{J_0, k}(x) + \sum_{j\geq J_0}^{\infty} \sum_{  k} d_{j,  k}\psi_{j,k}(x),
\ea
where $c_{J_0, k}$ indicates coarse coefficients, $d_{j,  k}$   detail coefficients, $\phi_{J_0,k}(x)$  scaling functions, and $\psi_{jk}(x)$  wavelet functions. We use different decomposing atom functions,  as scaling and wavelet functions, depending on a version of wavelet transform.
 For DWT, the atoms are
\ba
\phi_{J_0,k}(x)&=&2^{J_0/2}\phi(2^{J_0}x - k)\\
\psi_{jk}(x)&=&2^{j/2}\psi(2^jx - k),
\ea
where $x \in \mathbb{R}$, $j$ is a resolution level, $J_0$ is the coarsest resolution level, and $k$ is the location of an atom.

 For NDWT, atoms are
\ba
\phi_{J_0,k}(x)&=&2^{J_0/2}\phi(2^{J_0}(x - k))\\
\psi_{jk}(x)&=&2^{j/2}\psi(2^j(x - k)).
\ea
Notice that atoms in NDWT have a constant location shift $k$ at all levels, which yields the maximal sampling rate at each level. Two types of coefficients, $c_{J_0, k}$ and $d_{j,  k}$,   capture coarse and detail fluctuations  of an input signal, respectively. These are obtained as
\ba
c_{J_0,k} &=& \langle f(x), \phi_{J_0,k}  \rangle\\
d_{jk} &=& \langle f(x), \psi_{jk}  \rangle.
\ea
In a $p$-depth decomposition of an input signal of size $m$, a NDWT yields $m\times (p+1)$ wavelet coefficients, while DWT yields $m$ wavelet coefficients independent of $p$. The redundant transform NDWT decreases the variance of the scaling estimators, but at the same time, increases the correlations among wavelet coefficients.
  Since the estimators
of scaling are based on the second order properties of  wavelet coefficients, the NDWT-based estimators can be biased.

\begin{figure}[H]
\centering

	\includegraphics[width= 4.5in]{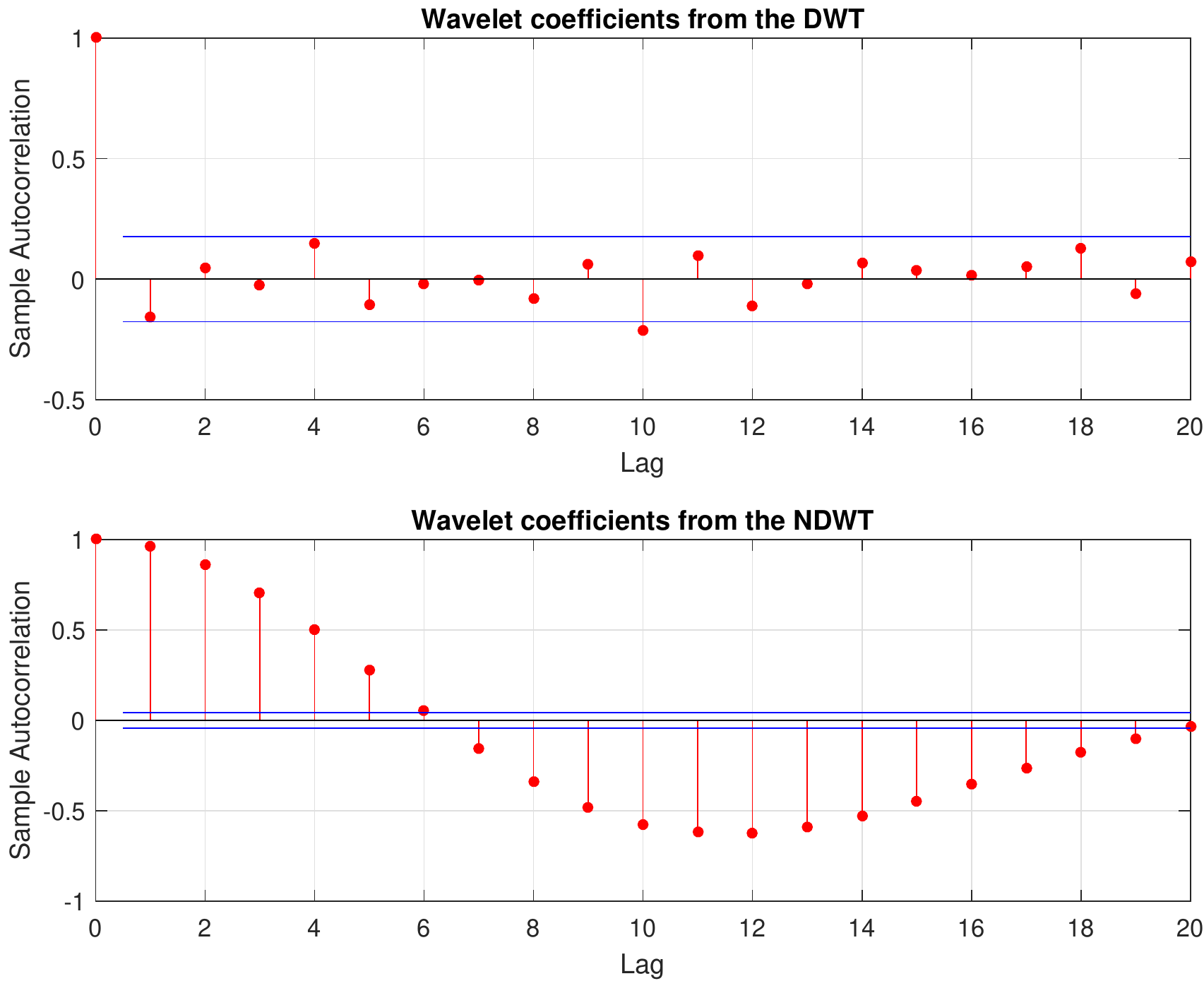}
\caption{The autocorrelation present in wavelet coefficients from the DWT and the NDWT.}
\label{fig:autocor}
\end{figure}
Figure \ref{fig:autocor} illustrates the autocorrelation within wavelet coefficients in the level $J-4$
(the level of finest detail is $J-1$, so $J-4$ is 4th ``most detailed'' level) in DWT and NDWT.   Haar wavelet was used on a Brownian motion path of size $2^{11}$. As we noted before, the coefficients from the NDWT are highly correlated while such correlation is not strong among the DWT  coefficients.

The two methods introduced in the following section reduce the effect of correlation among the coefficients,
while maintaining  redundancy and invariance as desirable threads of NDWT.

\section{ MEDL and MEDLA Methods}
We start by an overview of properties of wavelet coefficients and a brief literature overview methods in literature based on which we develop the proposed methods.

For defining a wavelet spectrum, and subsequently, for estimating $H$ only detail wavelet coefficients are used. When an fBm with Hurst exponent $H$ is mapped to the wavelet domain by DWT, the resulting detail wavelet coefficients satisfy the following properties \citep{tewfik1992correlation, abry1995wavelets, flandrin1992wavelet}:
\begin{enumerate}
\item[(i)]$d_{j}$, a detail wavelet coefficient at level $j$, follows the Gaussian distribution with  mean   0 and   variance  $\sigma_0^2 2^{-j(2H+1)}$, where $\sigma_0^2$ is the variance of a detail coefficient at level 0,
\item[(ii)] a sequence of wavelet coefficients from level $j$ is stationary, and
\item[(iii)] the covariance between two coefficients from any level of detail decreases exponentially as the distance between them increases;   the rate of the decrease depends additionally on the number of vanishing   moments of the decomposing wavelet.
\end{enumerate}
From property (i), the relationship between detail wavelet coefficients and Hurst exponent $H$ is
\begin{align*}
\log_2 \mathbb{E}\{d^2_j\}=-j(2H+1) + 2 \log_2 \sigma_0.
\end{align*}
\cite{abry2000wavelets} calculate sample variance of wavelet coefficients to estimate $\mathbb{E}\{d^2_j\}$ assuming i.i.d. Gaussianity of  coefficients at level $j$. Frequently, a squared wavelet coefficient is 
referred as an ``energy.''
 
Empirically, we look at the levelwise average of squared  coefficients (energies),
\begin{align*}
 \overline{d^2_j}=\frac{1}{n_j}\sum_{i=1}^{n_j}d^2_{j,k},
\end{align*}
where $n_j$ is the number of wavelet coefficients at level $j$.
The relationship between average energy  $\overline{d^2_j}$ and $H$ is
\begin{align*}
\log_2  \overline{d^2_j} \overset{d}{\approx} -(2H + 1 )j - \log_2 C - \log \chi^2_{n_j}/\log 2,
\end{align*}
where $\overset{d}{\approx}$ indicates approximate equality in distribution, $\chi^2_{n_j}$ follows a chi-square distribution with ${n_j}$ degrees of freedom, and $C$ is a constant. The method of \cite{abry2000wavelets} is affected by the non-normality of $\log_2  \overline{d^2_j}$ and correlation among detail wavelet coefficients, which results in biases of weighted least squares estimates. To reduce the bias, \cite{soltani2004estimation} defined ``mid-energies,''  as
\begin{align*}
D_{j,k}=\frac{d_{j,k}^2+d_{j,k+n_j/2}^2}{2}, k=1, ..., n_j/2.
\end{align*}
According to this approach, each multiresolution level is split on two equal parts and corresponding coefficients from each part are paired,
squared, and averaged. This produces a
quasi-decorrelation effect.
\cite{soltani2004estimation} show that level-wise averages of $\log_2 D_{j,k}$ are
asymptotically normal with the mean $-(2 H + 1)j + C,$ which is used to estimate $H$ by regression.

The estimators in \cite{soltani2004estimation} consistently outperform the estimators 
that use log-average energies, under various settings.  \cite{shen2007robust} show  that the method of \cite{soltani2004estimation} yields more accurate estimators since it takes the logarithm
of a mid-energy, and then averages. Moreover, averaging logged squared wavelet coefficients, rather than taking logarithm of averaged squared wavelet coefficients, is theoretically justified and this approach will be pursued in this paper. For both proposed methods,  MEDL and MEDLA, we first take the logarithm of a squared wavelet coefficient
 or an average of two squared wavelet coefficients, then derive the distribution of such logarithms under the assumption of independence. Next, we use the median of the derived distribution instead of the mean.
  The medians are more robust to potential outliers that can occur when logarithmic transform of a squared wavelet coefficient is taken and the magnitude of coefficient is close to zero. This
   numerical instability may increase the bias and variance of sample means. However, since the logarithms are monotone, the variability of the sample medians will not be affected.

The first proposed method is based on the relationship between the median of the logarithm of squared wavelet coefficients and the Hurst exponent. We use acronym ``MEDL'' to refer to this method. In MEDL, the logarithmic transform reduces the  autocorrelation, while the number of coefficients remains the same. The second method derives the relationship between the median of the logarithm of average of two squared wavelet coefficients and the Hurst exponent. We use acronym ``MEDLA'' to refer to this method. The MEDLA method is similar in concept to approach of \cite{soltani2004estimation} who paired
and averaged energies prior to taking logarithm. Then the mean of logarithms was connected to $H$.
 Instead, we repeatedly sample with replacement $m$ random pairs keeping distance between them at least $q_j$.
 Then, as in \cite{soltani2004estimation} we find the logarithm of pair's average and connect the Hurst exponent with the median of the logarithms.
As we relax the constraints on the distance between energies in each pair, we obtain a larger amount of distinct samples and selecting only $N$ samples out of such sample population further reduces the correlation.


 \begin{figure}[h]
\centering
\includegraphics[width= 4.5in]{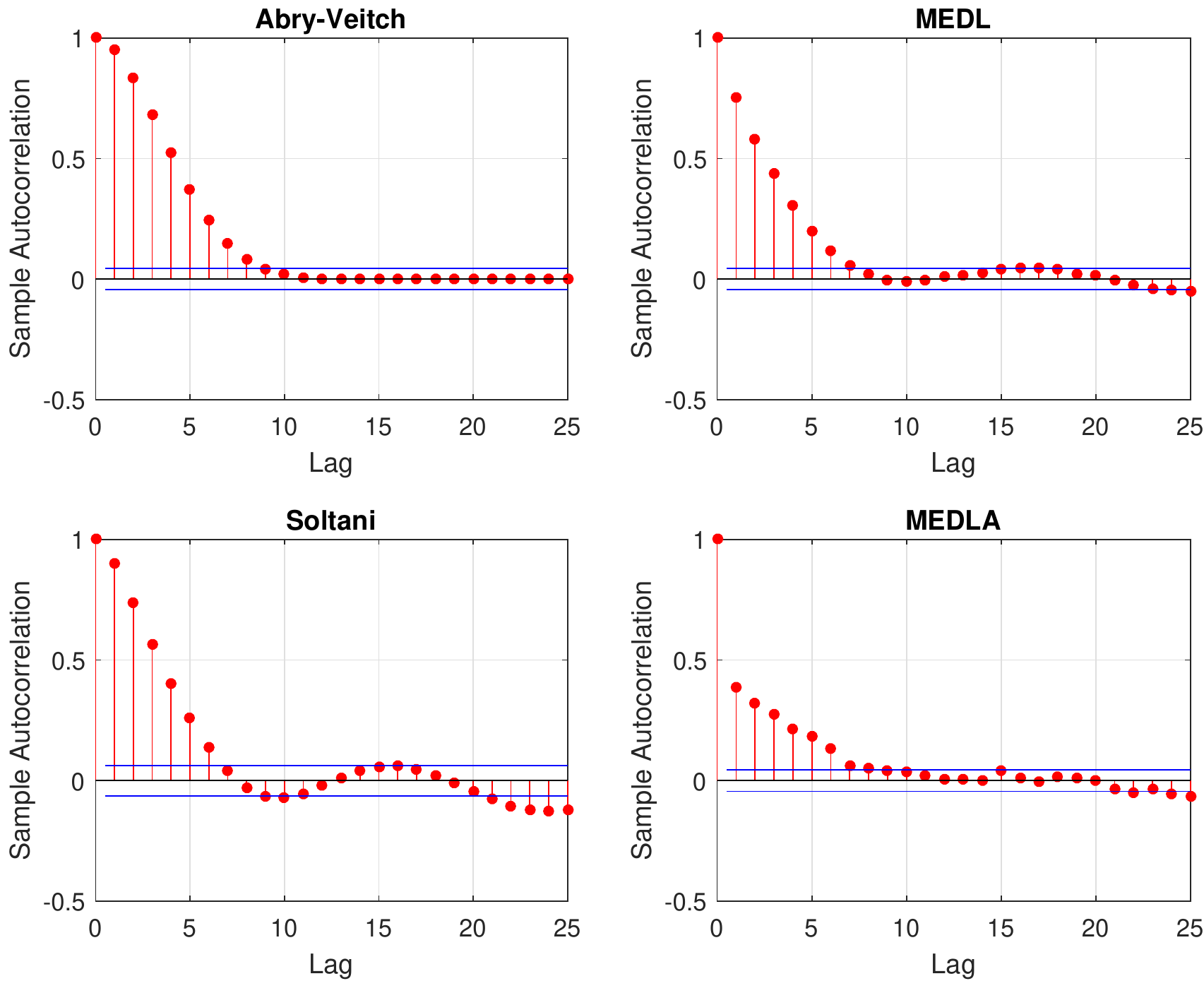} \caption{Autocorrelation of variables used in four methods.}
\label{fig:coeff}
\end{figure}
\noindent

To illustrate the decorrelation effects of the proposed methods, in Figure \ref{fig:coeff}, we compare the autocorrelation present in variables that are averaged: means of $d_{jk}^2$ for traditional method, means of $\log_2 \bigg[(d_{jk}^2+d_{j,k+m/2}^2)/2 \bigg]$  for Soltani-like method, medians of $\log d_{jk}^2$  for MEDL, and
medians of sampled $\log \bigg[(d_{jk_1}^2+d_{jk_2}^2)/2 \bigg]$ for MEDLA method. The two default methods exhibit higher amount of autocorrelation that decreases at a slower rate. The MEDLA shows substantial reduction in correlation.

For formal distributional assessment of the two proposed methods,
we start with an arbitrary wavelet coefficient from decomposition level $j$  at location $k$, $d_{jk}$, resulting from a non-decimated wavelet transform of a one-dimensional fBm $B_H(\omega, t), t \in \mathbb{R}$,
\ba
d_j=\int_{\mathbb{R}} B_H(\omega, t)\psi_{jk}(t)dt, \text{for some fixed } k.
\ea
As \cite{flandrin1992wavelet} showed, the distribution of a single wavelet coefficient is
\begin{align}
d_j  \stackrel{d}{=} 2^{-(H+1/2) j} \sigma Z,
\label{eq:wcoef}
\end{align}
where $Z$ follows a standard normal distribution, and $\sigma^2$ is the variance of wavelet coefficients at level 0. We will use (\ref{eq:wcoef}) repeatedly for the derivations that follow.


\subsection{MEDL Method}
\label{sec:MEDL}
For the median of the logarithm of squared wavelet coefficients (MEDL) method, we derive the relationship between the median of the logarithm on an arbitrary squared wavelet coefficient from decomposition level $j$ and Hurst exponent $H$. The following theorem serves as a basis for the MEDL estimator:
\begin{theorem}
\label{thm:MEDL}
Let $y_j^*$ be the median of $\log d_j^2$, where $d_j$ is an arbitrary wavelet coefficient from level $j$ in a NDWT of a fBm with Hurst exponent $H$.
 Then, the population median  is
 \be
 \label{th:MEDLy}
 y^*_j = - \log 2\, (2 H + 1) j + C,
 \ee
 where $C$ is a constant independent of $j$.
 The Hurst exponent can be estimated as
\be
\label{th:MEDLH}
\widehat{H}= - \frac{\widehat{\beta}}{2\log 2} -\frac{1}{2},
\ee
where $\widehat{\beta}$ is the slope in ordinary least squares (OLS) linear regression on pairs $(j,{\hat y}_j^*),$
and ${\hat y^*_j}$ is the sample median.
\end{theorem}
\noindent
The proof of Theorem \ref{thm:MEDL} is deferred to Appendix A. We estimate $y_j^*$ by taking sample median of logged energies at each level. The use of OLS linear regression is justified by the fact that variances of the sample medians ${\hat y^*_j}$ are constant in $j$, that is,
\begin{lemma}
\label{th:MEDLV}
The variance of sample median ${\hat y^*_j}$ at level $j$ is approximately
\ba
 \frac{\pi e^Q}{2 N  Q},
\ea
where $N$ is the sample size and $Q = \left( \Phi^{-1}(3/4) \right)^2$.
\end{lemma}
\noindent
The theorem is stating that the logarithm acts as a variance stabilizing operator;
the variance of the sample median is independent of level $j$, and ordinary regression to find slope $\beta$ in Theorem \ref{thm:MEDL} is fully justified.
Note that the use of OLS regression is not adequate in DWT; the weighted regression is needed to account for
levelwise heteroscedasticity.

\begin{figure*}
\begin{center}
        \subfigure{
            \includegraphics[width=0.45 \columnwidth]{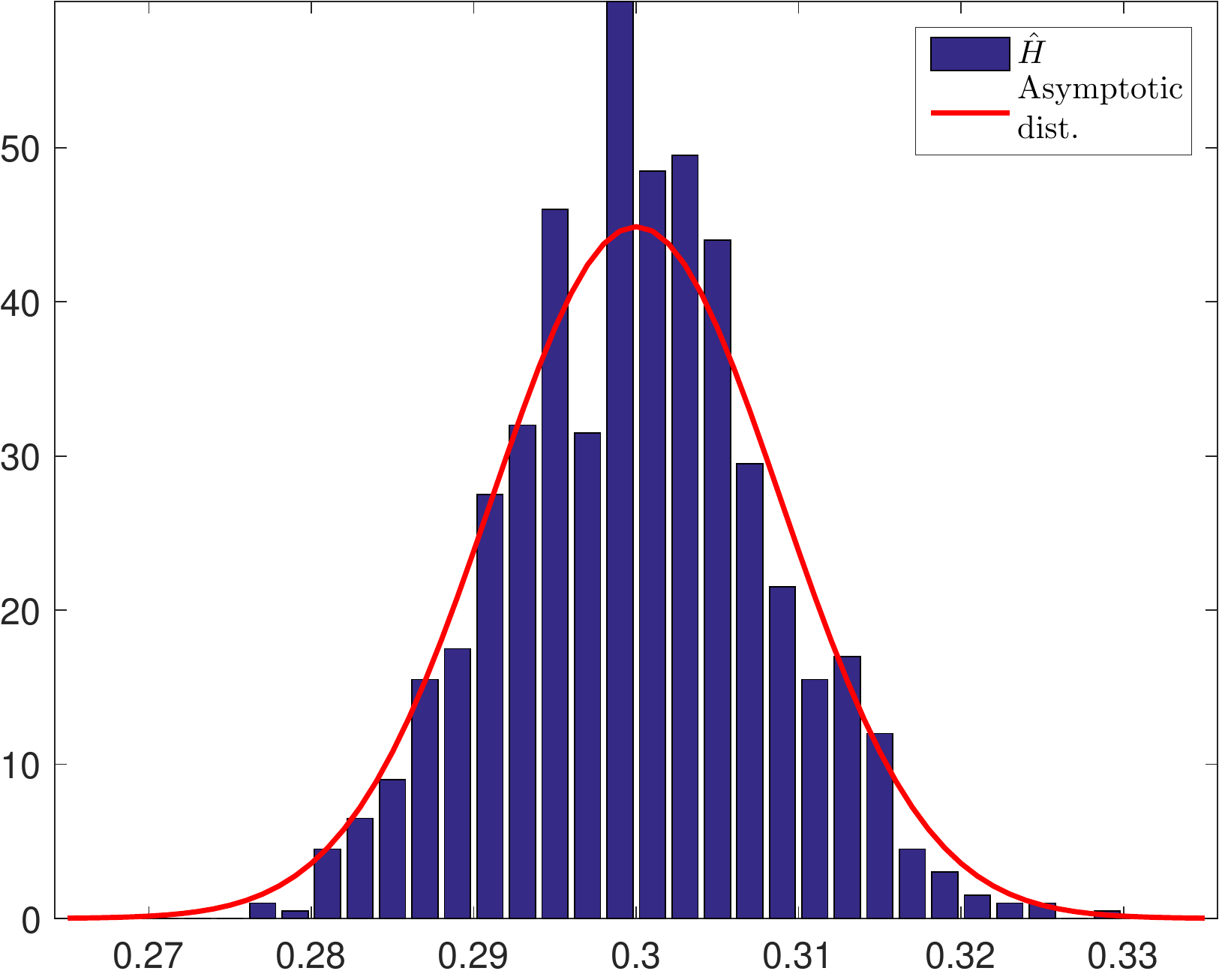}
        }
        \subfigure{
            \includegraphics[width=0.45 \columnwidth]{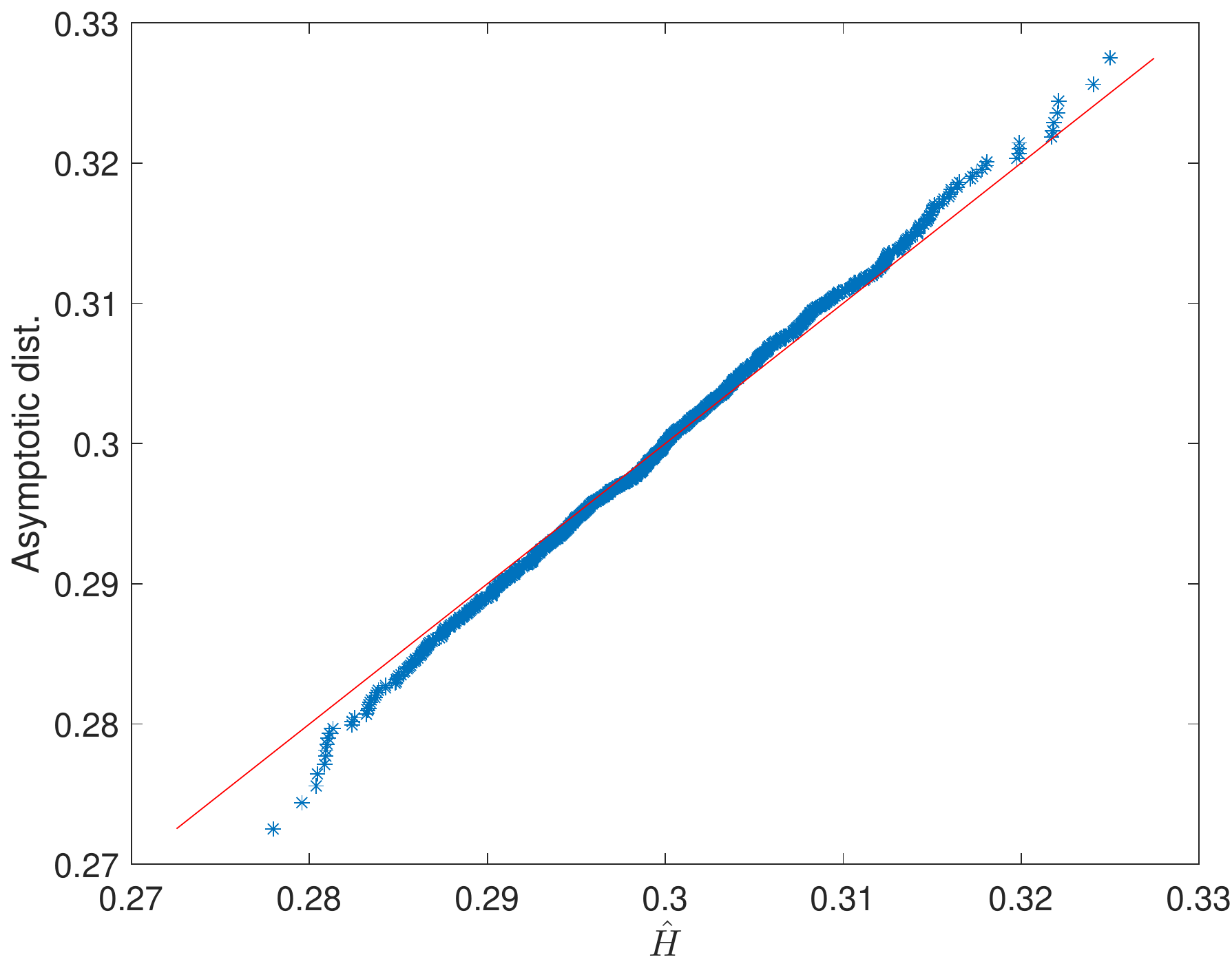}
        }
        \subfigure{
            \includegraphics[width=0.45 \columnwidth]{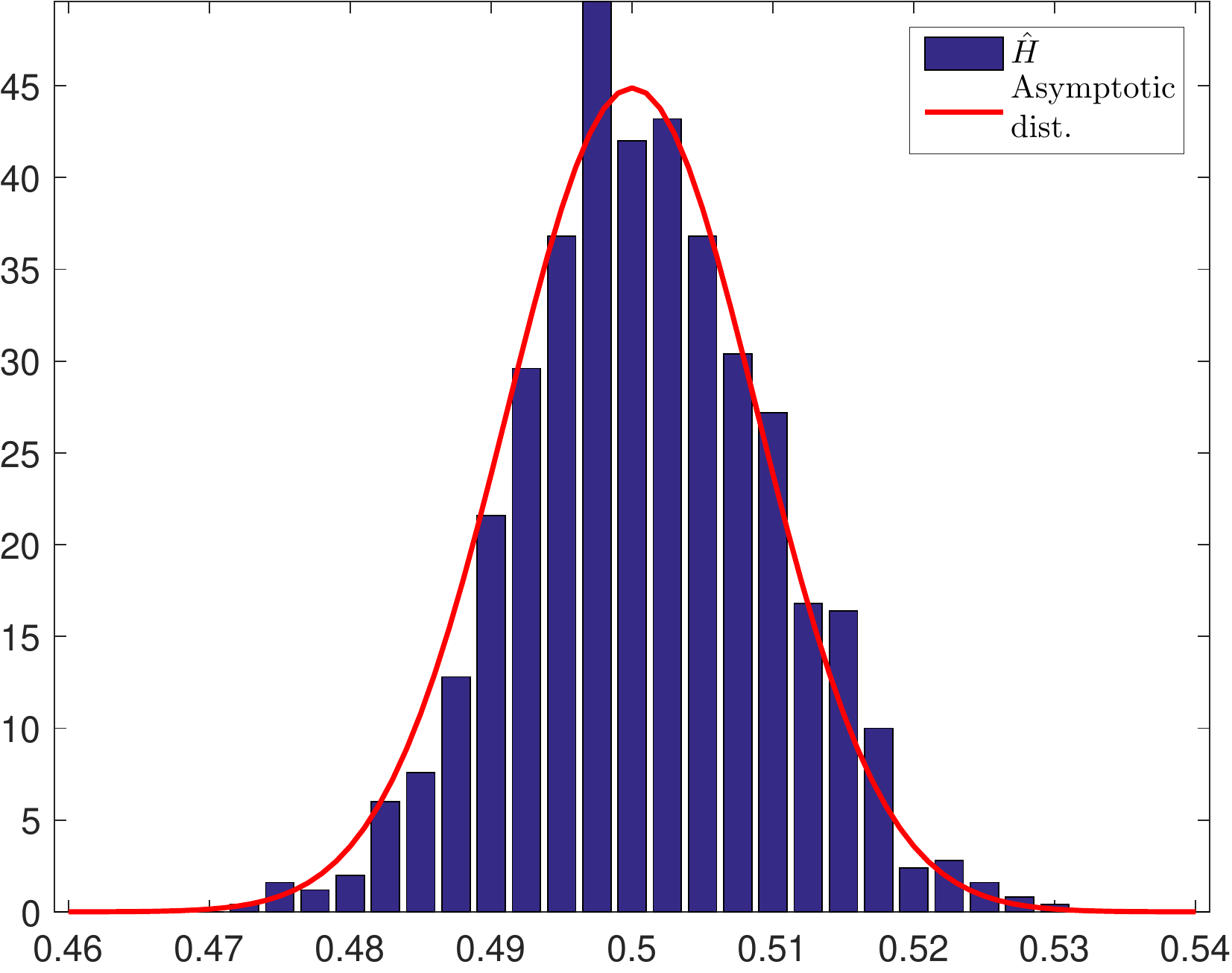}
        }
        \subfigure{
            \includegraphics[width=0.45 \columnwidth]{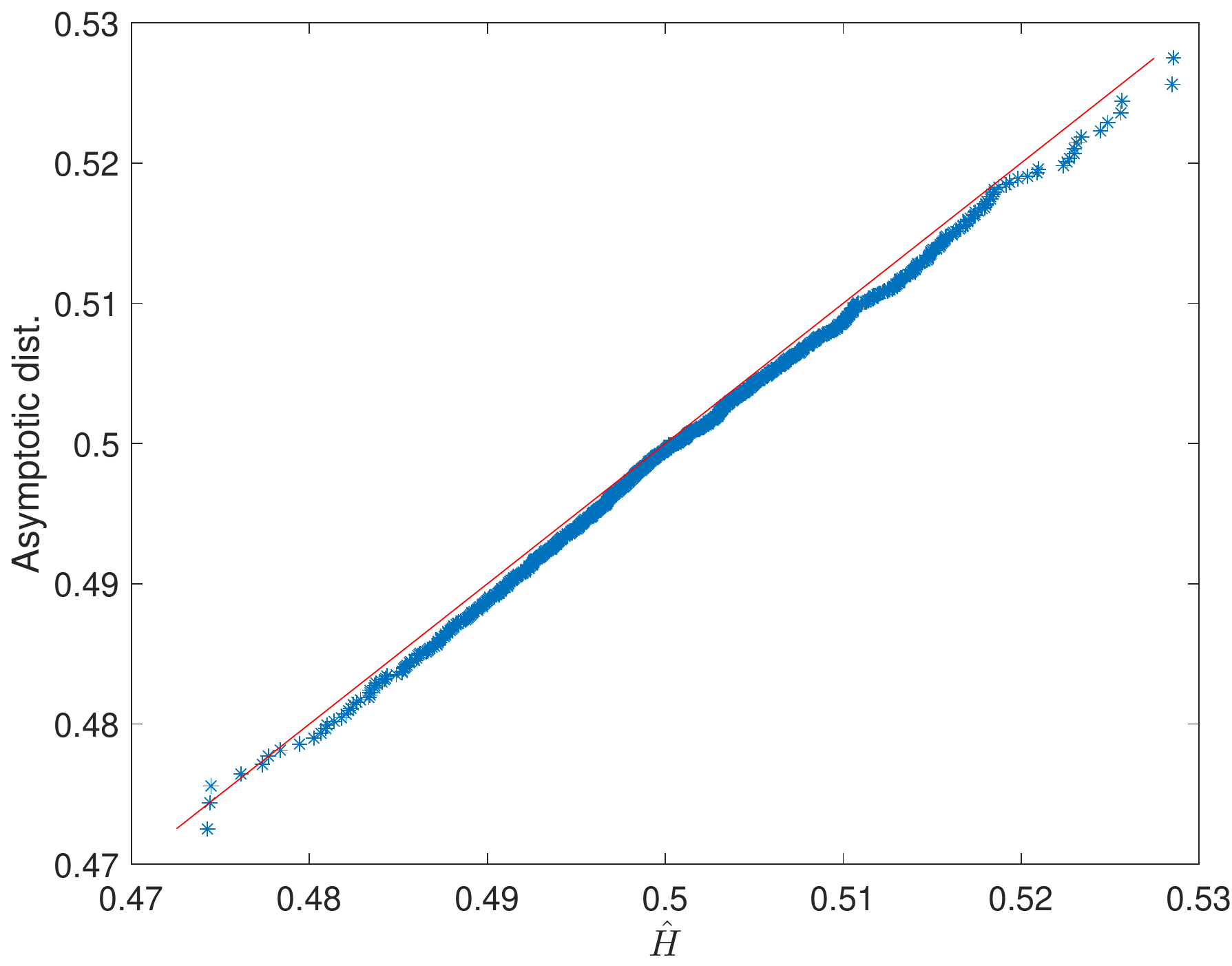}
        }
        \subfigure{
            \includegraphics[width=0.45 \columnwidth]{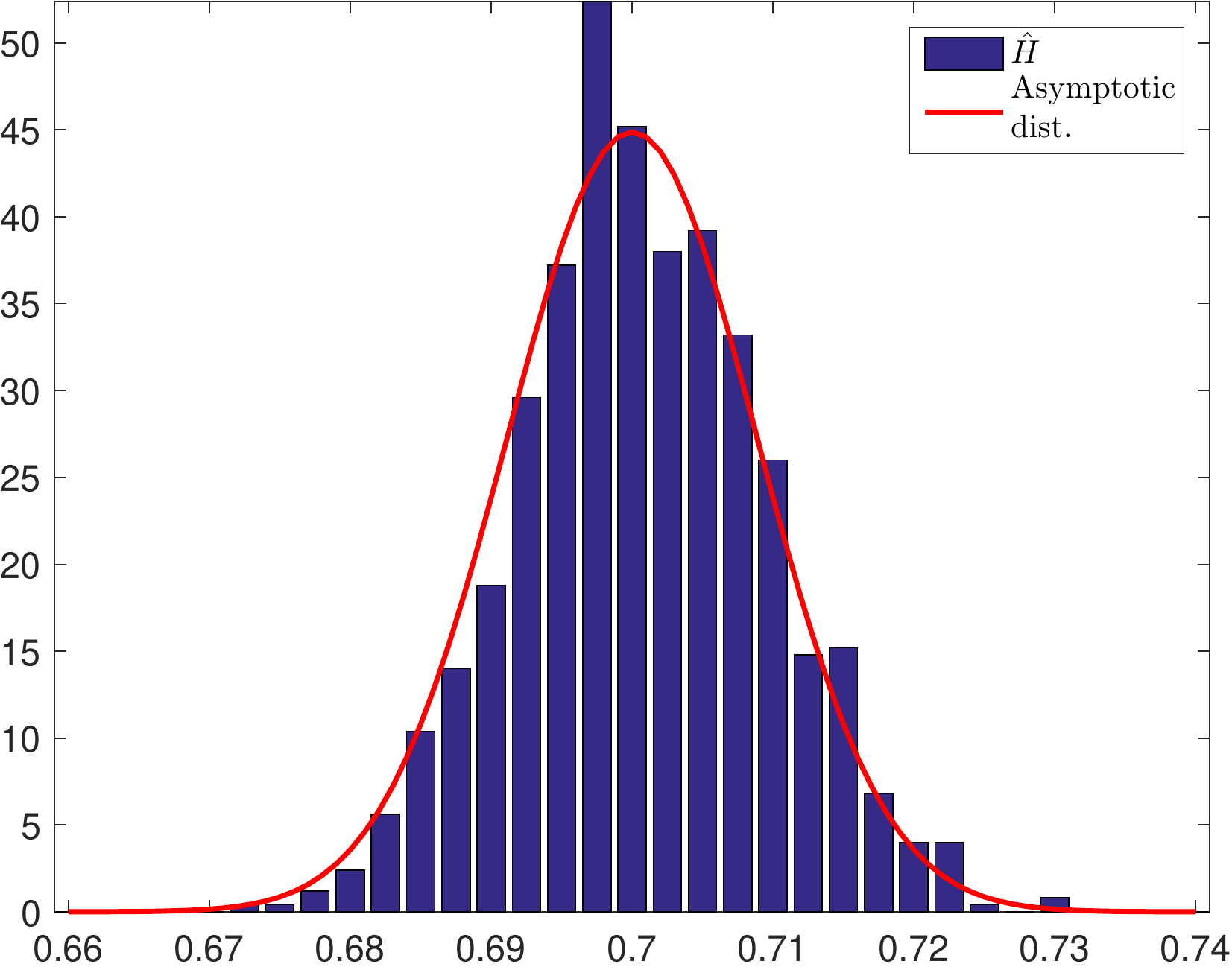}
        }
        \subfigure{
            \includegraphics[width=0.45 \columnwidth]{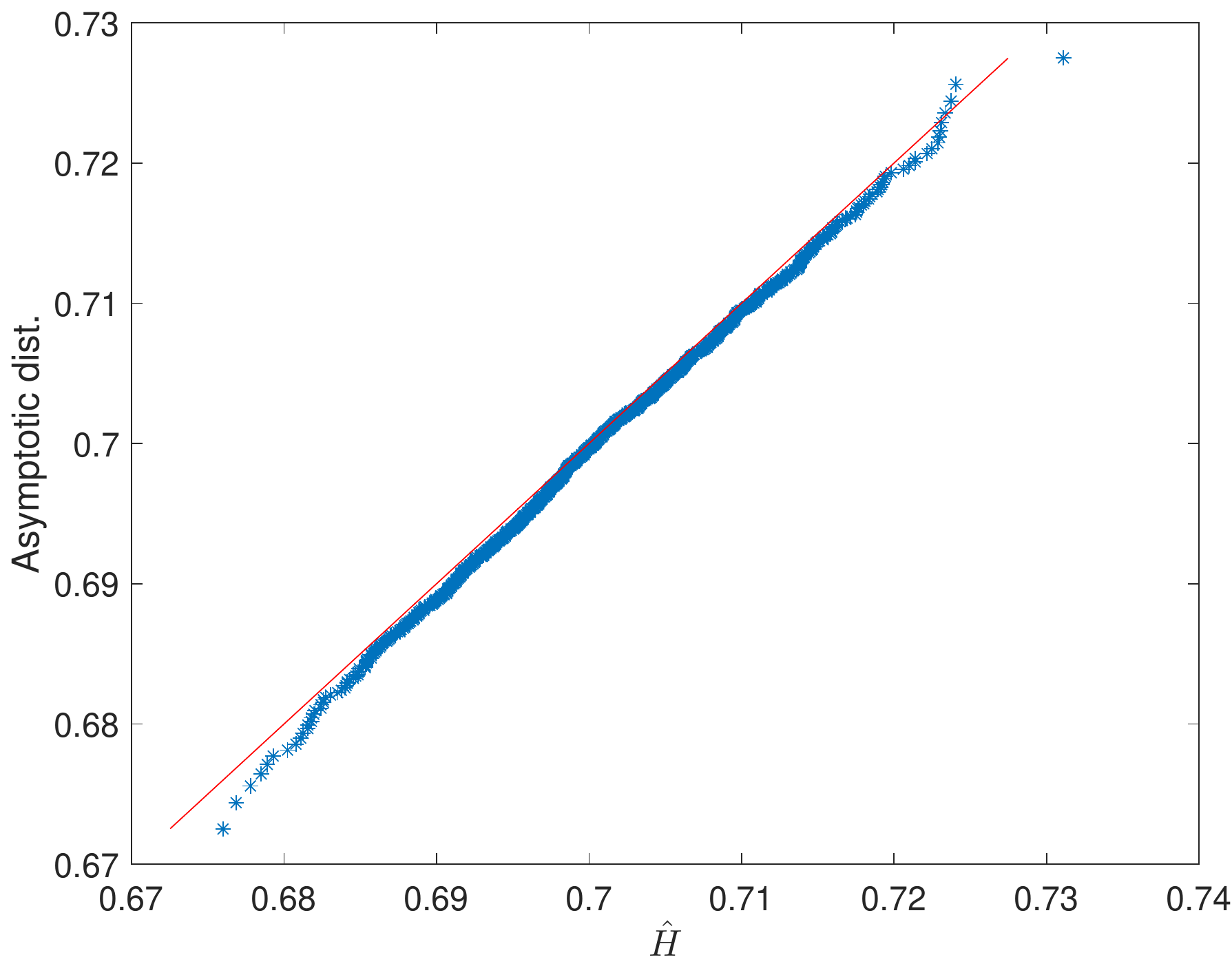}
        }
\end{center}
\caption{Panels on the right are histograms of $\widehat{H}$ and panels on the left are q-q plots of $\widehat{H}$ versus the quantiles asymptotic distribution when $H =0.3, 0.5,$ and 0.7, respectively.}
\label{fig:MEDLH}
\end{figure*}

The levelwise variance is approximately $ 5.4418/N,$ independent of $H$ and $\sigma^2.$
The proof of Theorem \ref{th:MEDLV} is deferred to Appendix A. In addition, for $\widehat{H}$ the normal approximation applies:
\begin{theorem}
The MEDL estimator $\widehat{H}$ follows the asymptotic normal distribution
\begin{align*}
\widehat{H} \overset{approx}{\sim} {\cal N} \left(H, \frac{3 A}{N m( m^2 - 1)(\log 2)^2}\right),
\end{align*}
\label{th:MEDLHD}
where $A =  \pi e^Q/(2 Q) \cong 5.4418$, $N$ is the sample size, and $m$ is the number of levels used in the spectrum.
\end{theorem}
\noindent
The proof of Theorem \ref{th:MEDLHD} is deferred to Appendix A. To illustrate Theorem \ref{th:MEDLHD}, we perform an NDWT of depth 10 on simulated fBm's with $H =0.3, 0.5,$ and $0.7$. We use resulting wavelet coefficients from levels $J-7$ to $J-2$
inclusive (i.e., six levels) to estimate $H$ with MEDL. Following Theorem \ref{th:MEDLHD}, $\widehat{H}$ of MEDL in the simulation follows a normal distribution with mean $H$ and variance $7.9007\times 10^{-5}$, which is illustrated in Figure \ref{fig:MEDLH}.

\subsection{MEDLA Method}
\begin{figure*}
\begin{center}
        \subfigure{
            \includegraphics[width=0.45 \columnwidth]{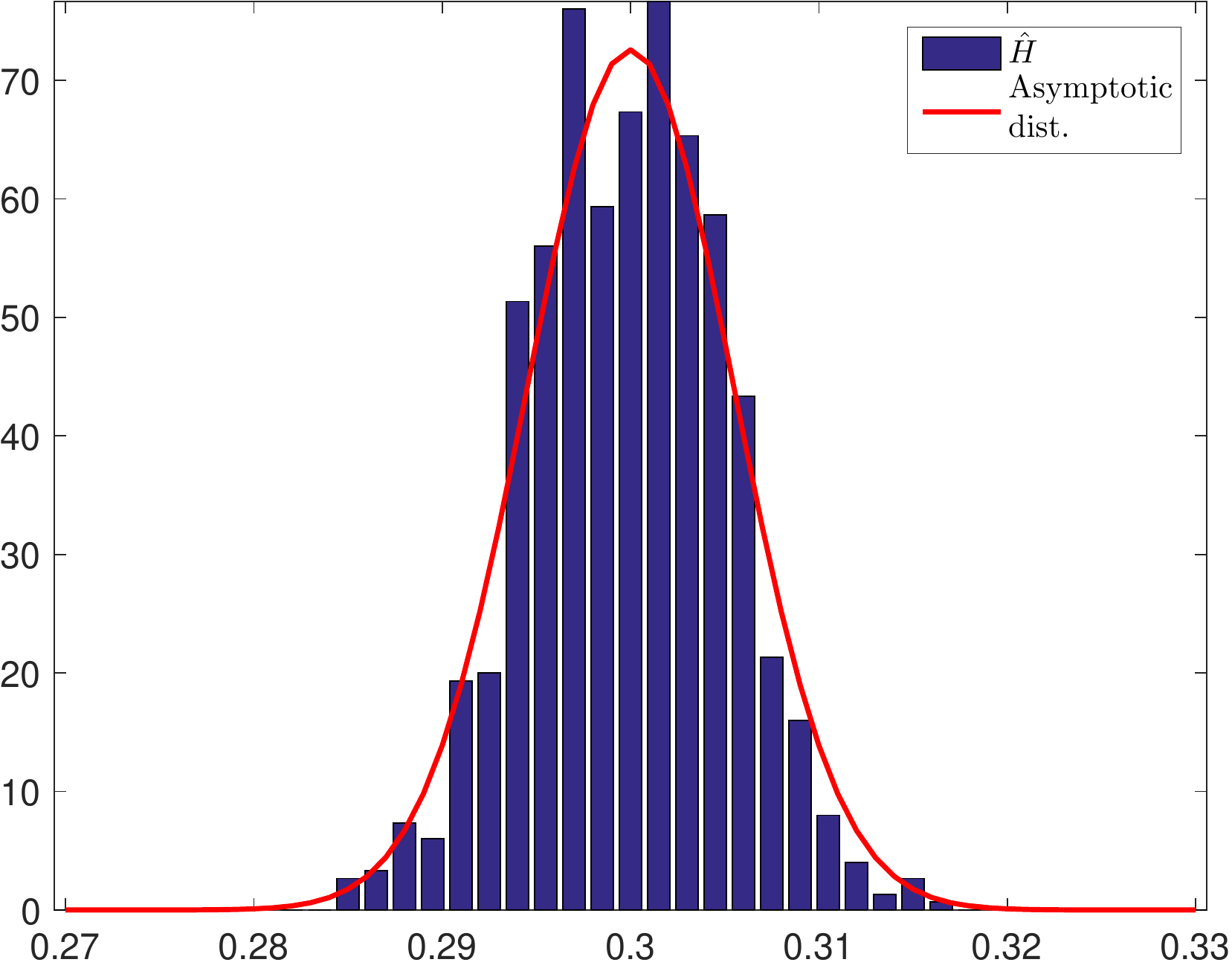}
        }
        \subfigure{
            \includegraphics[width=0.45 \columnwidth]{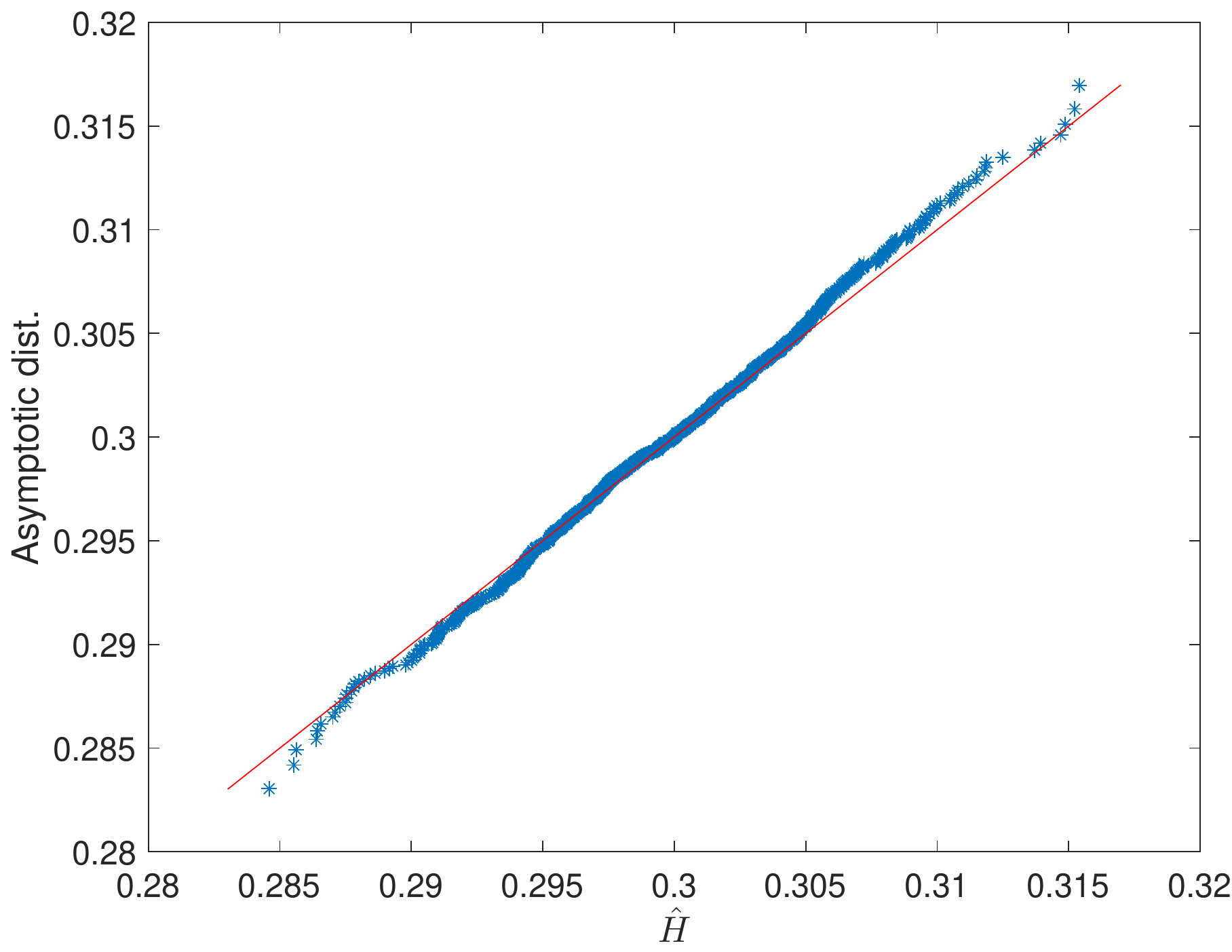}
        }
        \subfigure{
            \includegraphics[width=0.45 \columnwidth]{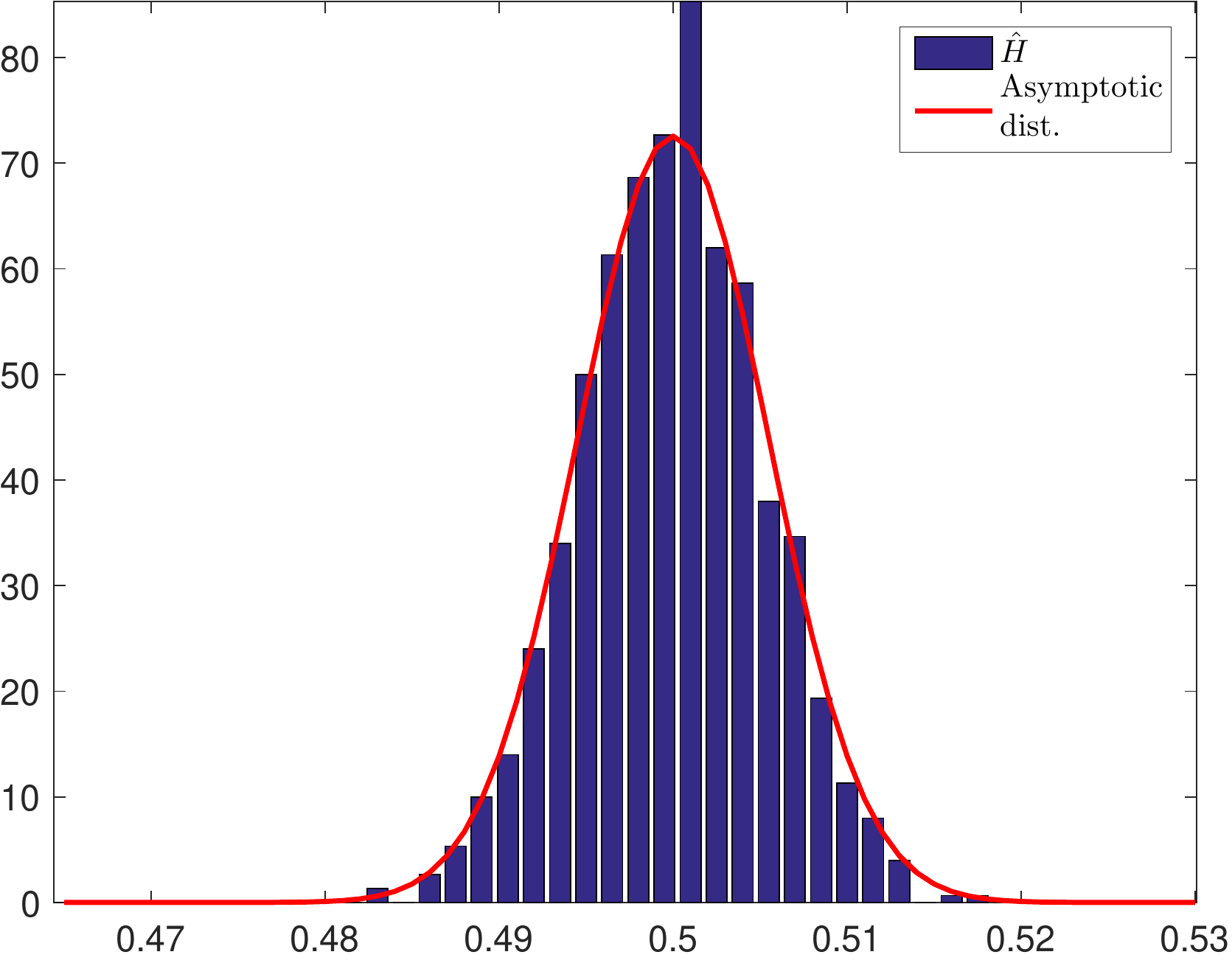}
        }
        \subfigure{
            \includegraphics[width=0.45 \columnwidth]{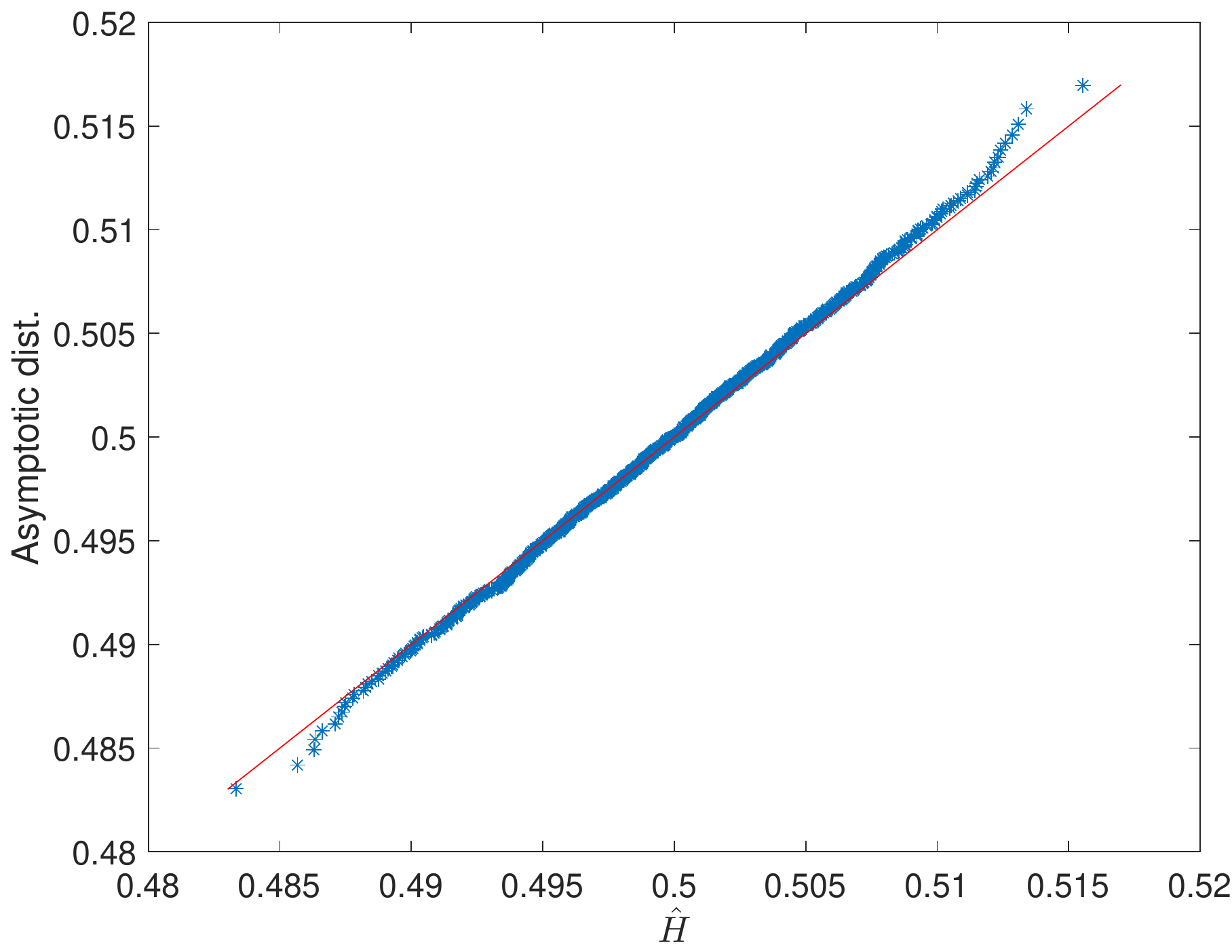}
        }
        \subfigure{
            \includegraphics[width=0.45 \columnwidth]{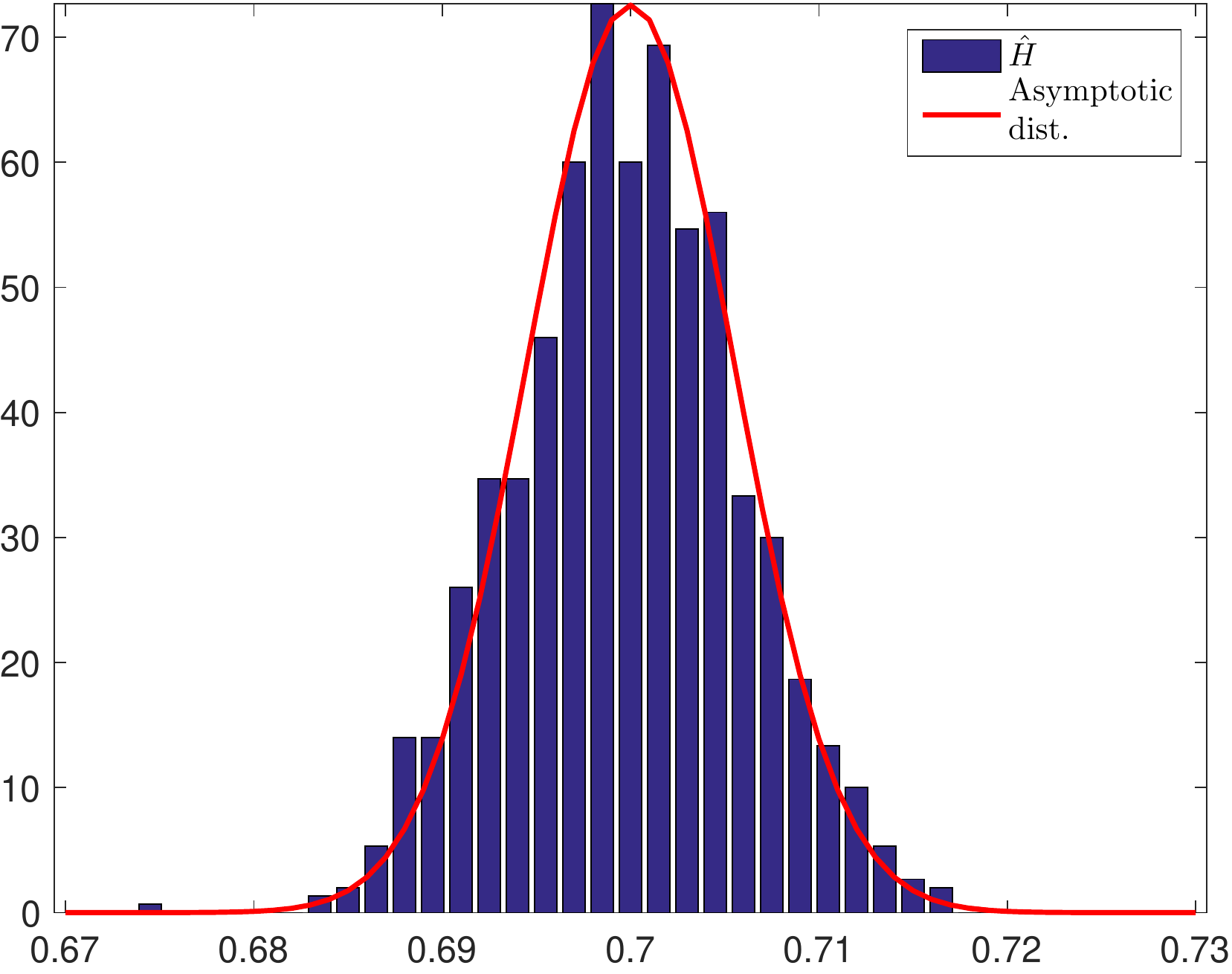}
        }
        \subfigure{
            \includegraphics[width=0.45 \columnwidth]{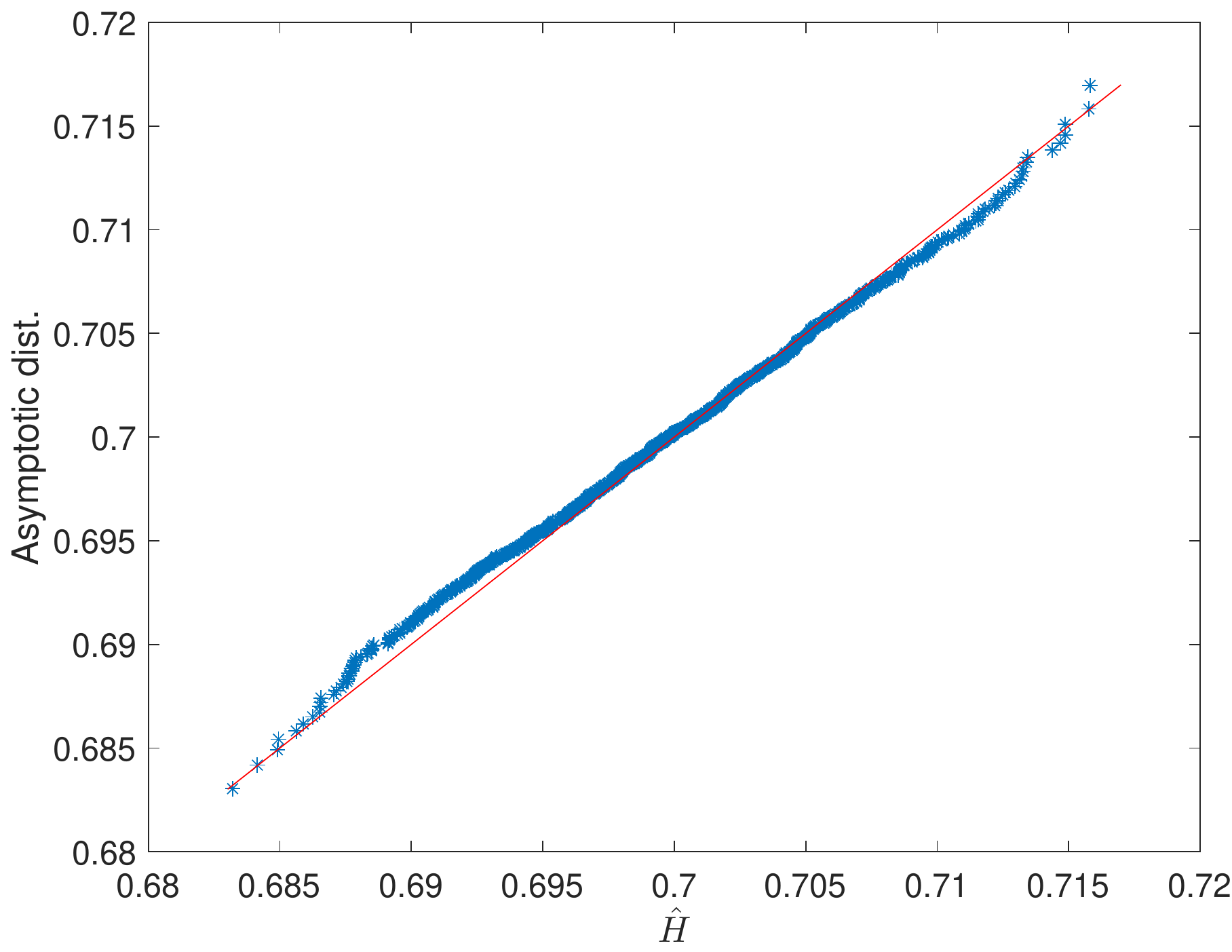}
        }
\end{center}
\caption{Panels on the right are histograms of $\widehat{H}$ and panels on the left are q-q plots of $\widehat{H}$ versus the quantiles of asymptotic distribution when $H =0.3, 0.5,$ and 0.7, respectively.}
\label{fig:MEDLAH}
\end{figure*}

For the median of the logarithm of averaged squared wavelet coefficients (MEDLA) method, we derive the relationship between logarithm of an average of two energies and Hurst exponent $H$. \cite{soltani2004estimation} proposed a method that quasi-decorrelates wavelet coefficients by splitting all wavelet coefficients from one level into left and right sections and pairing every coefficient in the left section with its counterpart in the right section, maintaining an equal distance to its pair (i.e., members in each pair are $m/2$ apart when $m$ is the number of  wavelet coefficients on that level). Then, \cite{soltani2004estimation} averaged every pair of energies and took logarithm of each average. We follow similar idea except that instead of fixing the combinations of pairs, which amounts to $m/2$ pairs in \cite{soltani2004estimation}, we randomly sample with replacement $m$ pairs whose members are at least $q_j$ apart. Based on sample autocorrelation graphs, we define $q_j=2^{J-j}$ that decrease with level $j$ because the finer the subspace (i.e., larger $j$), the lower the correlation among wavelet coefficients. Then, we propose an estimator of $H$ based on the following result.
\begin{theorem}
\label{thm:MEDLA}
Let $d_{jk_1}$ and $d_{jk_2}$ be  two  wavelet coefficients from level $j$, at positions $k_1$ and $k_2$, respectively, from a NDWT of a fBm
with Hurst exponent $H$. Assume that $|k_1 - k_2| > q_j,$ where $q_j$ is the minimum separation distance
that depends on level $j$ and selected wavelet base.
Let $y_j^*$ be the median of $\log \bigg[\frac{d_{jk_1}^2+d_{jk_2}^2}{2}\bigg]$. Then,
as in Theorem \ref{thm:MEDL}, results
(\ref{th:MEDLy}) and (\ref{th:MEDLH}) hold.
\end{theorem}
\noindent
The proof of Theorem \ref{thm:MEDLA} is deferred to Appendix B. To estimate $y_j^*$, we first repeatedly sample $m$ pairs of wavelet coefficients with replacement from all pairs that are at least $q_j$  apart. Then, we take logarithm of pair's average energy and take the median. As in Theorem \ref{th:MEDLV}, the variances of sample medians ${\hat y^*_j}$ are free of $j$.
\begin{lemma}
\label{th:MEDLAV}
The variance of the sample median $\hat{y^*_j}$ at level $j$ is approximated by
\begin{align*}
\frac{1}{N (\log  2)^2},
\end{align*}
where $N$ is the sample size.
\end{lemma}
\noindent
The proof is straightforward and given in Appendix B. Thus, the variance of ${\hat y^*_j}$ is constant over levels. We find that MEDLA estimator of $H$ indeed follows a approximately normal distribution with a mean and a variance given in the following theorem.
\begin{theorem}
The estimator $\widehat{H}$ of MEDLA follows the asymptotic normal distribution
\begin{align*}
\widehat{H} \overset{approx}{\sim} {\cal N}\left(H, \frac{3}{ N m( m^2 - 1)(\log 2)^4} \right),
\end{align*}
\label{th:MEDLAHD}
where $N$ is the sample size, and $m$ is the number of levels used in the spectrum.
\end{theorem}
The proof of Theorem \ref{th:MEDLAHD} is deferred to Appendix B. To illustrate Theorem \ref{th:MEDLAHD}, we use the same wavelet coefficients from the simulation in section \ref{sec:MEDL}. Following Theorem \ref{th:MEDLAHD}, $\widehat{H}$ of MEDLA in the simulation follows an approximate normal distribution with mean $H$ and variance $7.9007\times 10^{-5}$, which is shown in Figure \ref{fig:MEDLAH}.

\section{Simulations}
\begin{figure*}
\begin{center}
        \subfigure[$H$=0.3]{
            \includegraphics[width=0.52 \columnwidth]{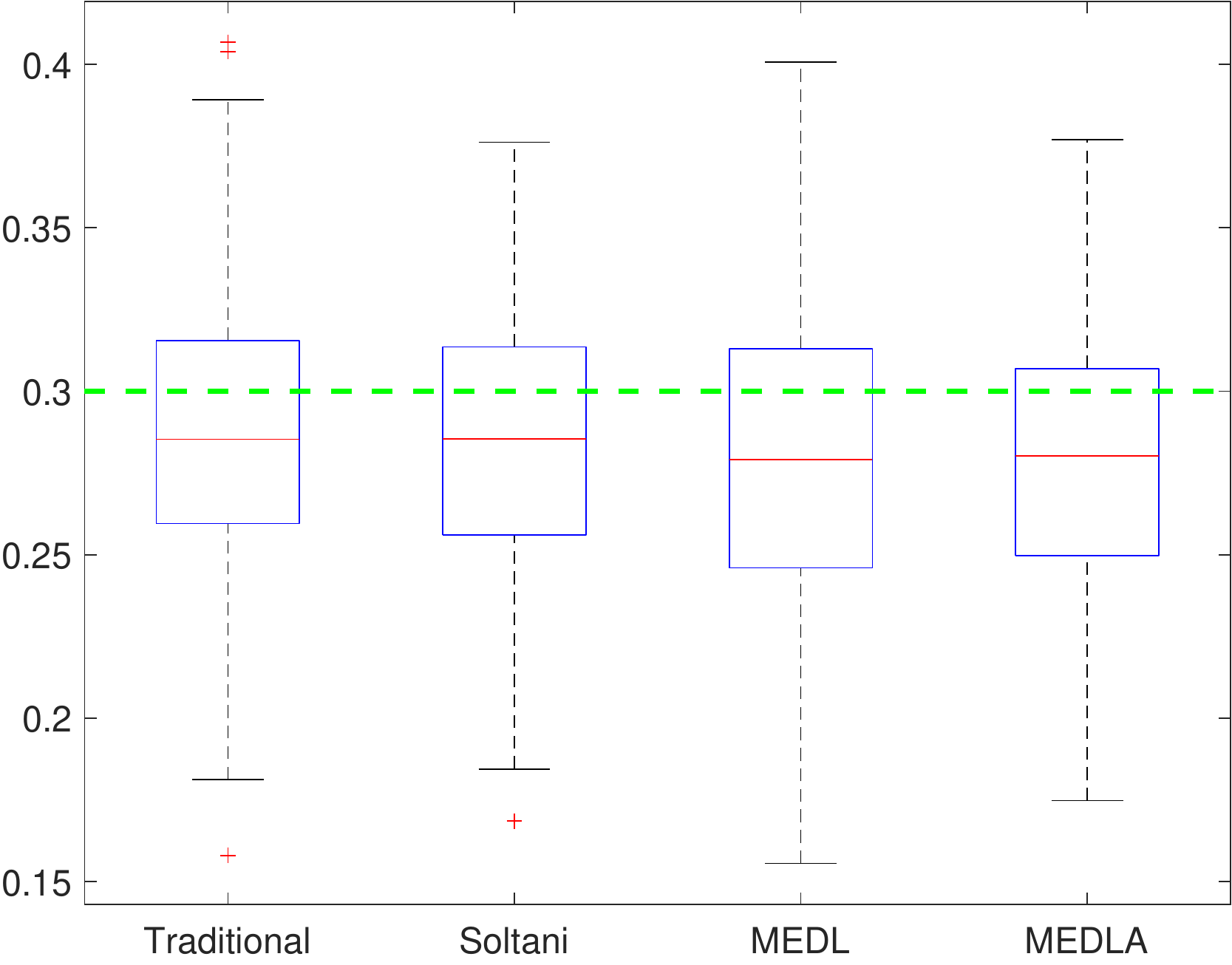}
        }
        \subfigure[$H$=0.5]{
            \includegraphics[width=0.52 \columnwidth]{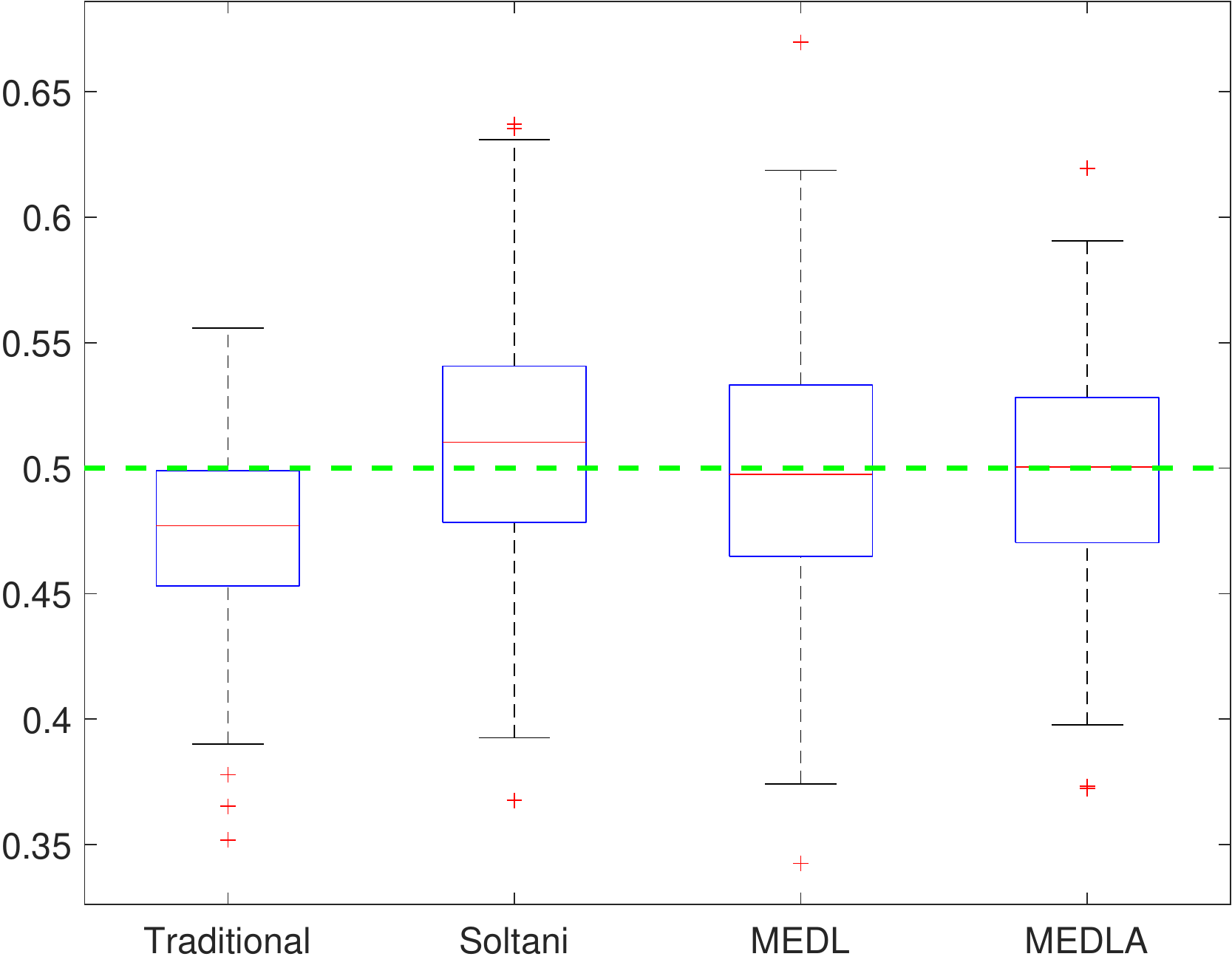}
        }
        \subfigure[$H$=0.7]{
            \includegraphics[width=0.52 \columnwidth]{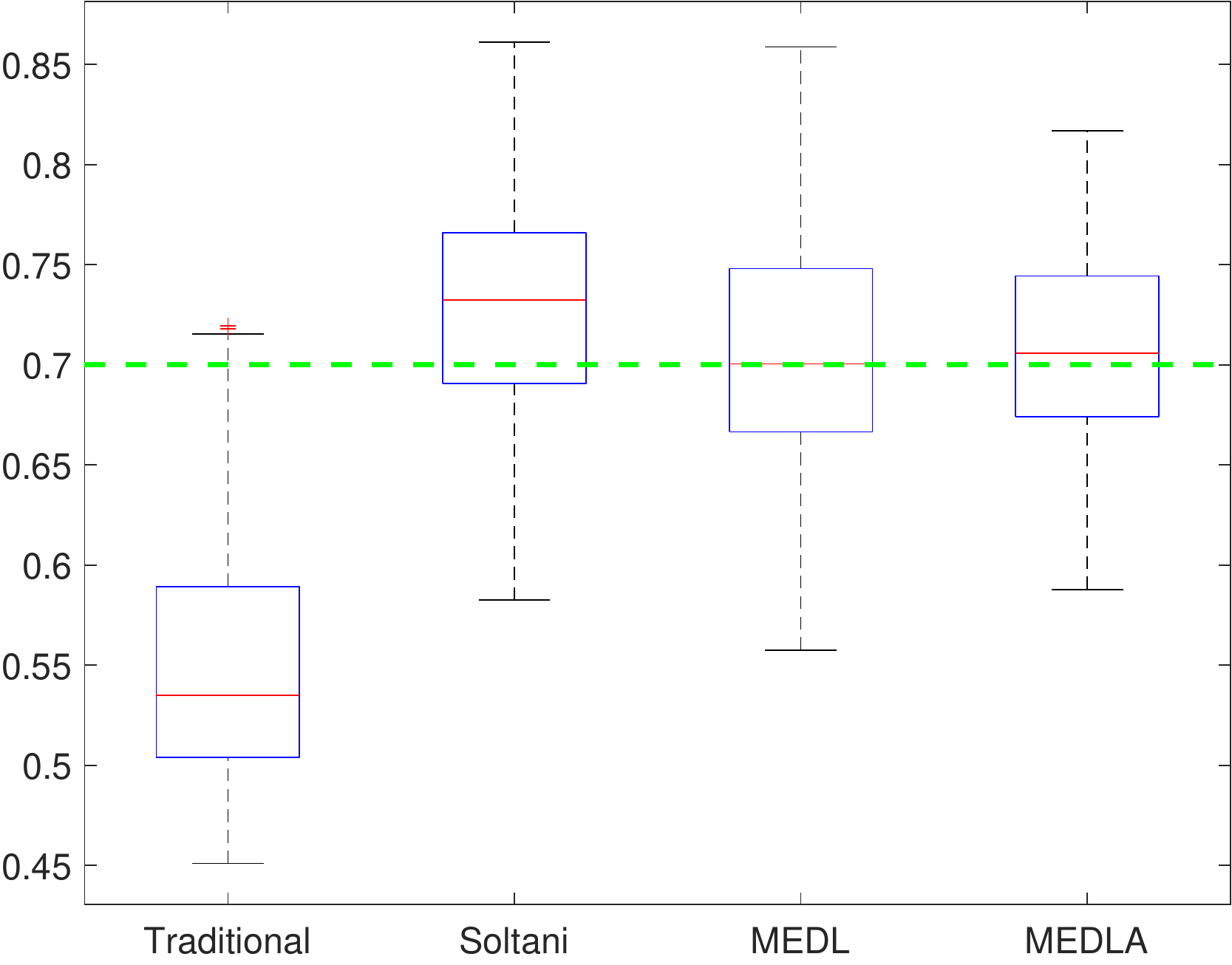}
        }
\end{center}
\caption{Boxplots of $\widehat{H}$ by four methods with 300 simulated 1-D fBm signals of size $2^{11}$ when $H$=0.3, 0.5, and 0.7}
\label{fig:MEDLA_simul}
\end{figure*}

Next, we assess the performance of MeDL and MEDLA in estimation of Hurst exponent,
We simulate three sets of three hundred one-dimensional fractional Brownian motion (1-D fBm) paths of size $2^{11}$ with Hurst exponents 0.3, 0.5, and 0.7 respectively. Then, we perform an NDWT of depth 10 with a Haar wavelet on each simulated signal and obtain wavelet coefficients to which we apply MEDL and MEDLA. 
For all methods and estimations, we use wavelet coefficients from levels $J-7$ to $J-2$ in the regression. We compare the estimation performance of the proposed methods to two standard methods: a method of  \cite{veitch1999wavelet} and a method of \cite{soltani2004estimation}, both in the context of NDWT. We present the estimation performance in terms of mean, variance, bias-squared, and mean squared error, based on 300 simulations for each case. Table \ref{tb:sim_stat} and Figure \ref{fig:MEDLA_simul} indicate that as $H$ increases, the proposed methods outperform the standard methods. For smaller $H$, the estimation performance of all methods is comparable.

\begin{table}[h]
\begin{center}
\begin{tabular}{| c| c| c |c | c |}
\multicolumn{1}{c}{$H$=0.3} & \multicolumn{1}{c}{} & \multicolumn{1}{c}{ } & \multicolumn{1}{c}{} & \multicolumn{1}{c}{ } \\ \hline
Method & 	Traditional & 	Soltani &  MEDL & 	MEDLA\\ \hline
Mean & 	0.2864 & 	0.2849 & 	0.2778  & 	0.2783\\ \hline
Variance & 	0.0017 & 	0.0015& 	0.0021 & 	0.0016  \\ \hline
Bias-squared & 	0.0002 & 	0.0003 &  	0.0005&	0.0005  \\ \hline  MSE & 	0.0019 & 	0.0018 &  	0.0026&	0.0021 \\ \hline
\multicolumn{1}{c}{ } & \multicolumn{1}{c}{} & \multicolumn{1}{c}{ } & \multicolumn{1}{c}{} & \multicolumn{1}{c}{ } \\
\multicolumn{1}{c}{$H$=0.5} & \multicolumn{1}{c}{} & \multicolumn{1}{c}{ } & \multicolumn{1}{c}{} & \multicolumn{1}{c}{ } \\ \hline
Method & 	Traditional & 	Soltani &  MEDL & 	MEDLA\\ \hline
Mean & 	0.475 & 	0.5091 & 	0.4966 & 	0.4982 \\ \hline
Variance & 	0.0012 & 	0.0022  & 	0.0023 & 	0.0017\\ \hline
Bias-squared & 	0.0006 & 	6.7E-5  & 4.1E-6 & 	1.3E-6\\ \hline
MSE & 	0.0018 & 	0.0023 & 	0.0023 	& 0.0017  \\ \hline
\multicolumn{1}{c}{ } & \multicolumn{1}{c}{} & \multicolumn{1}{c}{ } & \multicolumn{1}{c}{} & \multicolumn{1}{c}{ } \\
\multicolumn{1}{c}{$H$=0.7} & \multicolumn{1}{c}{} & \multicolumn{1}{c}{ } & \multicolumn{1}{c}{} & \multicolumn{1}{c}{ } \\ \hline
Method & 	Traditional & 	Soltani &  MEDL & 	MEDLA\\ \hline
Mean & 	0.5524 & 	0.7286  & 	0.7065 & 	0.7084 \\ \hline
Variance & 	0.0039 & 	0.0028 & 	0.0033 & 	0.0024 \\ \hline
Bias-squared & 	0.0217 & 	0.0008 & 	3.3E-5& 	6.2E-5  \\ \hline
MSE & 	0.0256 & 	0.0036 & 	0.0033 & 	0.0024 \\ \hline
\end{tabular}
\caption{Estimation of $H$ with 300 simulated 1-D fBm signals of size $2^{11}$ when $H$=0.3, 0.5, and 0.7 by four methods}
\label{tb:sim_stat}
\end{center}
\end{table}

\section{Conclusions}
We proposed two methods for robust estimation of Hurst exponent in one- and two-dimensional signals that scale. Unlike the standard methods, the proposed methods are based on NDWT. The motivation for using NDWT was its
redundancy and time-invariance. However, the redundancy, which was useful for
the stability of estimation, increases autocorrelations among the wavelet coefficients.
The proposed methods lower the present autocorrelation by (i) taking logarithm of the squared wavelet coefficients
prior to averaging, (ii) relating the Hurst exponent to the median of the model distribution, rather than the mean,
and (iii) resampling the coefficients.

The methods are compared to  standard approaches and give estimators with smaller MSE for a range of
input conditions.

Instead of medians in (ii) we could employ any other quantile; the methodology is equivalent and will differ for
the intercept and variance in the regressions.

\section*{References}
\bibliography{BS}

\section*{Appendix}

\subsection*{A. Derivation of MEDL}
\label{ap:MEDL}
{\bf Proof of Theorem \ref{thm:MEDL}}\\
A single wavelet coefficient in a non-decimated wavelet transform of fBm(H) is normally distributed, with variance depending on
its level $j$,
\ba
d_j & \overset{d}{=} & {\cal N}(0, 2^{-(2H+1)j} \sigma^2  ).
\ea
Its rescaled energy is $\chi^2$ with one degree of freedom,
\ba
\delta = \frac{2^{(2H+1)j}}{\sigma^2}d_j^2  \stackrel{d}{=}   \chi_1^2,
\ea
with density
\ba
 \frac{\delta^{1/2-1} (\frac{1}{2})^{1/2}}{\Gamma(\frac{1}{2})}e^{-\delta/2}
= \frac{e^{-\delta/2}}{ \sqrt{2\delta}\Gamma(\frac{1}{2}) }.
\ea
The pdf of $d_j^2$ is
\ba
f(d_j^2)=\frac{e^{-c_jd_j^2/2}}{\sqrt{2c_jd_j^2}\Gamma(1/2)}c_j,
\ea
where $c_j=\frac{2^{(2H+1)j}}{\sigma^2}$.
Let $y=\log d_j^2$, then $d_j^2=e^y$ and $\frac{\partial d_j^2}{\partial y}=e^y$. The pdf of $y$ is
\ba
f(y)&=&\frac{c_je^{\frac{-c_je^y}{2}}}{\sqrt{2c_je^y}\Gamma(1/2)} e^y
=\frac{\sqrt{c_j}e^{-\frac{c_je^y}{2}}e^{y/2}}{\sqrt{2}\Gamma(1/2)} = \sqrt{\frac{c_j}{2\pi}}e^{-\frac{c_j e^y}{2}}e^{y/2},
\ea
 The cdf of $y$ is
\ba
F(y)= \int_{-\infty}^{y}f(t) dt =   2\Phi\left(\sqrt{c_j} e^{y/2} \right)-1,
\ea
where $\Phi$ is the cdf of standard normal distribution.
Let $y^*$   be the median of the distribution of $y$. We obtain the expression of $y*$ by solving
$F(y^*) = 1/2$. This results in
\ba
y^*  = 2\log \left[ \frac{1}{\sqrt{c_j}} \Phi^{-1}(3/4)\right]
\ea
From this equation, we can find a link between $y^*$ and the Hurst exponent $H$ by substituting $c_j,$
\be
\label{eq:med}
y^*&=& 2\log [\Phi^{-1}(3/4)] -\log c_j  \nonumber \\
&=& - \log 2 \,(2H+1)j  + \log \sigma^2 + 2\log [\Phi^{-1}(3/4)] \\
&=&  - \log 2 \,(2H+1)j + C, \nonumber
\ee
where $C$ is a constant independent on the level $j$.

{\bf Proof of Lemma \ref{th:MEDLV}}\\
An approximation of variance of sample median ${\hat y}_j^*$ is obtained
using normal approximation to a quantile of absolutely continuous distributions,
\ba
Var( {\hat y}_j^* ) \approx  \frac{1}{4 N (f(y^*_j))^2 }.
\ea
After substituting the expression for $y^*$
we obtain Lemma \ref{th:MEDLV}
\ba
Var( \hat y_j^* ) \approx  \frac{\pi e^Q}{2 N Q},~~~~~Q= \left[\Phi^{-1}(3/4)\right]^2 \approx 0.4549.
\ea
Thus the variance of the sample median is approximately $5.4418/N.$\\

{\bf Proof of Theorem \ref{th:MEDLHD}}\\
An NDWT-based spectrum uses the pairs
\ba
\left(j, ~\hat{y}^*_j \right), ~j = J-m-a-1, \dots, J-a-1.
\ea
from $m$ decomposition levels, starting with a coarse $j=J-m-a$ level and ending with finer level
$j=J-1-a$. Here $a$ is an arbitrary integer between 0 and $J-3$. When $a=0$, the finest level $j=J-1$ until level $J-1-m$ are used.

Then the spectral slope is
\ba
\hat \beta = \frac{12}{ m (m^2 - 1)}  \sum_{j=J-m-a-1}^{J-1-a} (j - J -  a -(m+1)/2)  ~ {\hat y}^*_j.
\ea

The estimator $\widehat \beta$ is unbiased,
\ba
E \hat \beta &=& \frac{12}{  m (m^2 - 1)}  \sum_{j=J-m-a-1}^{J-1-a} (j -J - a - (m+1)/2) ~ \left( -\log 2\, (2 H + 1)j + C \right) \\
&=&  -\log 2\, (2 H + 1),
\ea
where $C$ is a constant and $H$ is the theoretical Hurst exponent.

By substituting $\mbox{Var}( {\hat y}^*_j) = A/N$ from Theorem \ref{th:MEDLV}
we find
\ba
\mbox{Var}(\widehat \beta) = \frac{12 A}{N m (m^2 - 1)}, \mbox{  and  ~~}
\mbox{Var}(\widehat H ) =  \frac{3 A}{N m( m^2 - 1)(\log 2)^2},
\ea
for $\widehat H =-{\widehat \beta}/{ (2 \log 2)}-1/2.$

Thus, the MEDL estimator $\widehat H$ is approximately normal with mean $H$ and variance $3 A/(N m( m^2 - 1)(\log 2)^2),$
 where $A \cong 5.4418$, $N$ is the sample size, and $m$ is the number of levels used for the spectrum.

\subsection*{B. Derivation of MEDLA} \label{ap:MEDLA}
{\bf Proof of Theorem \ref{thm:MEDLA}}.\\
We begin by selecting the pair of wavelet coefficients that follow a normal distribution with a zero mean and a variance dependent on level $j$, from which the wavelet coefficients are sampled.
\ba
d_{j,k_1}, d_{j,k_2} & \sim & {\cal N}( 0, 2^{-(2H+1) j} \sigma^2 ),
\ea
where $\sigma$ is the standard deviation of wavelet coefficients from level 0, $k_1$ and $k_2$ are  positions of wavelet coefficients in level $j$, and $H$ is the Hurst exponent. We also assume that coefficients $d_{j,k_1}$ and $d_{j,k_2}$ are independent, which is a reasonable assumption when
the distance $|k_1 - k_2|>q_j=2^{J-j}$.
Then, we define $\delta$ as
\ba
\delta &= &\frac{2^{(2H+1)j}}{\sigma^2}(d_{j,k_1}^2+d_{j,k_2}^2)   = C_j \cdot a,
\ea
for $C_j = \frac{2 \cdot 2^{(2H+1)j}}{\sigma^2} \text{ and }\;  a=\frac{d^2_1+d^2_2}{2} $.
Since $\delta$ follow $\chi^2_2$ distribution, the pdf of the average of two squared wavelet coefficients $a$ is
\ba
f(a) & = & \frac{C_j}{2}e^{-\frac{C_ja}{2}}.
\ea
Denote $y = \log a$. The pdf and cdf of $y$ are
\ba
f(y) & = & \frac{C_j}{2}e^{-\frac{C_je^y}{2}}e^y\\
F(y) & = & 1-e^{-C_je^y/2}.
\ea
Let $y^*$ be  the median of $y$. The expression for $y^*$ is obtained by
solving
$F(y^*)  = 1-e^{-C_je^{y^*}/2}=1/2,$
\ba
y^* = \log\,(\log 4) - \log C_j.
\ea

After replacing $C_j$ with $\frac{2 \cdot 2^{(2H+1)j}}{\sigma^2},$ the median becomes
\ba
 y^* &= -\log 2 \,(2H+1) j + \log \sigma^2 + \log\,(\log 2),
\ea
similarly as in (\ref{eq:med}) in the MEDL method.\\

{\bf Proof of Lemma \ref{th:MEDLAV}}\\
An approximation of variance of sample median is obtained as
\ba
Var( \hat y_j^* ) \approx  \frac{1}{4 n (f(y^*_j))^2 }.
\ea
After plugging in the expression for $y^*$ into $\frac{1}{4 n (f(y^*_j))^2 }$,
we obtain
\ba
Var( \hat y_j^* ) \approx  \frac{1}{ N (\log 2)^2 },
\ea
Thus the variance of the sample median in MEDLA method is approximately $2.08/N.$

{\bf Proof of Theorem \ref{th:MEDLAHD}}\\
For the distribution of $\widehat{H}$ from MEDLA, we follow the same regression steps on pair $(j,\hat{y_j^*})$ as in Appendix A. By substituting $\mbox{Var}( {\hat y}^*_j) =\frac{1}{N (\log 2)^2}$ from (\ref{th:MEDLAV}) we find
\ba
\mbox{Var}(\widehat \beta) = \frac{12}{N m (m^2 - 1)  (\log 2)^2 }, \mbox{  and  ~~}
\mbox{Var}(\widehat H ) =  \frac{3}{ N m( m^2 - 1)  (\log 2)^4},
\ea
for $\widehat H =-{\hat \beta}/{ (2 \log 2})-1/2.$ Thus, the MEDLA estimator $\hat H$ is approximately normal with mean $H$ and variance $3/( N m( m^2 - 1)(\log 2)^2),$ where $N$ is the sample size, and $m$ is the number of levels used for the spectrum.

\end{document}